\documentclass[a4paper]{article}
\usepackage[a4paper,includemp=false,body={16cm,24.7cm},vcentering]{geometry}
\usepackage[latin1]{inputenc}
\usepackage[T1]{fontenc}
\usepackage{mathptmx}
\usepackage[bf,center]{caption}
\usepackage{graphicx}
\usepackage{url,amsmath,amssymb,amsthm}

\makeatletter
\renewcommand{\section}{\@startsection{section}{1}{0mm}{30pt}{12pt}{\normalfont\normalsize\bfseries}}

\renewcommand{\subsection}{\@startsection{subsection}{2}{0mm}{18pt}{12pt}{\normalfont\normalsize\itshape}}
\makeatother

\newcommand{\Title}[1]{\begin{center}{\bfseries\fontsize{12pt}{12pt}\selectfont#1}\end{center}}
\newcommand{\Author}[2]{\begin{center}{\fontsize{12pt}{12pt}\selectfont#1}\\{\it #2~}\end{center}}
\newcommand{\Introduction}{\section*{Introduction}}

\newcommand{\Teff}{\ensuremath{T_{\rm eff}}}                      

\begin{document}

\Title{Eclipsing Binary Stars: the Royal Road to Stellar Astrophysics}

\Author{John Southworth}{Astrophysics Group, Keele University, Staffordshire, ST5 5BG, UK}


\Introduction

\noindent
Russell \cite{Russell48} famously described eclipses as the ``royal road'' to stellar astrophysics (see also \cite{Batten05}). From photometric and spectroscopic observations it is possible to measure the masses and radii (to 1\% or better!), and thus surface gravities and mean densities, of stars in eclipsing binary systems (EBs) using nothing more than geometry. Adding an effective temperature subsequently yields luminosity and then distance (or {\it vice versa}) to high precision. This wealth of directly measurable quantities makes EBs the primary source of empirical information on the properties of stars, and therefore a cornerstone of stellar astrophysics.

In this review paper I summarise the current standing of EB research, present an overview of useful analysis techniques, and conclude with a glance to the future. For a deeper discussion I recommend the peerless reviews by Popper \cite{Popper67} \cite{Popper80}, Andersen \cite{Andersen91} and Torres et al.\  \cite{Torres10}, and the textbook by Ron Hilditch \cite{Hilditch01}.


\section{The Ghost of Christmas Past}

\noindent
John Goodricke (ably supported by his assistant Edward Pigott) was in 1783 the first to invoke the concept of stellar eclipses, in order to explain the variations in the light of $\beta$\,Persei (Algol). William Herschel \cite{Herschel02} christened the term ``binary star'' in the course of his work on resolved astrometric binaries. In 1889, Edward Pickering triggered the study of close binary systems when he noticed that the star Mizar exhibited spectral lines which not only moved in wavelength but were sometimes doubled. Carl Vogel \cite{Vogel90} proved the binary nature of $\beta$\,Persei by detecting spectral line shifts before and after an eclipse, giving rise to the term ``spectroscopic binary''. Finally, Stebbins \cite{Stebbins11} discovered the first EB with a double-lined spectrum, $\beta$\,Aurigae (Menkalinan). His measurements of the masses and radii of the two component stars are surprisingly close to the modern values \cite{Me07}.

Whilst it is comparatively straightforward to interpret astrometric and radial velocity measurements of binary systems, their eclipses are more challenging. Henry Norris Russell \cite{Russell12a,Russell12b} provided the first rigorous mathematical structure for their comprehension and progressively developed the method before encapsulating it in a textbook \cite{Russell52}. Zden\v{e}k Kopal expounded a similarly encyclopaedic knowledge in his own textbook \cite{Kopal59}. His methods differed greatly from Russell's approach, and the two men were not on good terms \cite{Batten05}. The late 1960s and early 1970s saw the production of the first computer programs to model binary systems, which have revolutionised the study of eclipses by laying them on a proper physical basis. Chief architects of the revolution were Wilson and Devinney \cite{Wilson71}, the flexibility of whose code arises from the inclusion of full Roche geometry. An honourable (but slightly biased) mention goes to Paul Etzel \cite{Etzel75,Etzel81}, whose {\sc ebop} code is orders of magnitude faster but restricted to systems whose stars are well-separated and thus only modestly distorted. A large number and variety of other codes exist, although few are widely used and many are now of only historical interest.

\begin{figure}
\includegraphics[width=0.50\textwidth,angle=0]{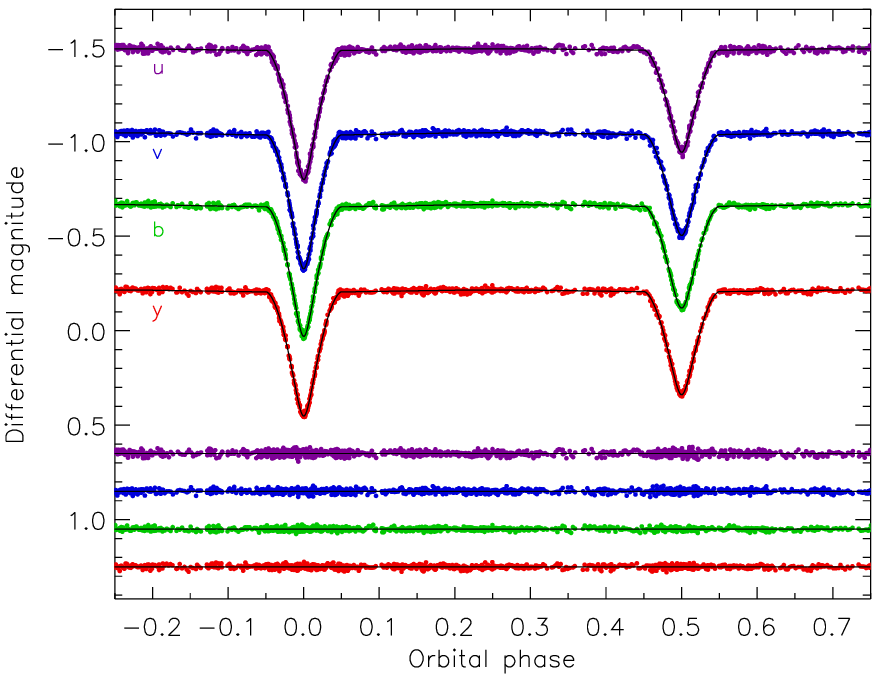}
\includegraphics[width=0.50\textwidth,angle=0]{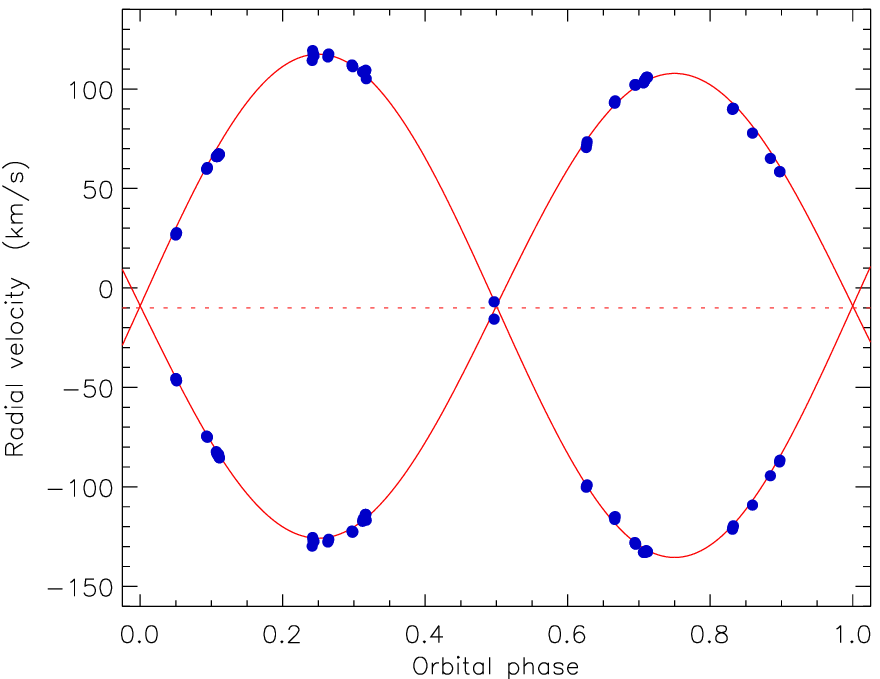}
\caption{Example observational data, for the dEB WW\,Aurigae \cite{Me05}.
{\it Left panel:} light curves in the Str\"omgren $uvby$ passbands plotted
as a function of orbital phase. The black lines show the best-fitting model
generated with the {\sc jktebop} code. The residuals of the fit are plotted
at the base of the figure, offset from zero for clarity. {\it Right panel:}
RV curves for WW\,Aur. The measured RVs are shown with blue circles and the
fitted curves with red lines.} \end{figure}

\section{The Ghost of Christmas Present}

\noindent
From this point I concentrate on detached eclipsing binaries (dEBs), whose component stars are not undergoing mass transfer except for the interception of stellar winds. These objects evolve as if they were two single stars, so are the most appropriate reference points for stellar astrophysics (specifically theoretical models of stellar evolution). The lack of mass transfer also makes them generally more tractable, although semi-detached (Algol) systems can be rather straightforward and also better distance indicators \cite{Wilson08}. A full study of one dEB requires both photometric and spectroscopic observations, covering a large fraction of its orbital phases.

\subsection{Photometric analysis of eclipsing binaries}

\noindent
The detailed shape of the two eclipses is related to the physical properties of the stars. The primary eclipse occurs, by definition, when the hotter star is behind the cooler star (superior conjunction). The primary eclipse is therefore deeper than the secondary eclipse (inferior conjunction). Exceptions to this rule are possible if the orbit is not circular, in which case different surface areas can be eclipsed during primary and secondary minimum thus altering their relative depth (or making one vanish entirely e.g.\ SW\,CMa \cite{Torres11}).

The most important quantities which can be constrained are the fractional radii of the two stars, defined as the actual radii divided by the orbital semi-major axis:
\begin{equation}
r_{\rm A} = \frac{R_{\rm A}}{a}
\qquad \qquad \qquad \qquad
r_{\rm B} = \frac{R_{\rm B}}{a}
\end{equation}
$r_{\rm A}$ and $r_{\rm b}$ can be quite correlated, so in practice it is often convenient to recast them as their sum and ratio:
\begin{equation}
r_{\rm A} + r_{\rm b}
\qquad \qquad \qquad \qquad
k = \frac{r_{\rm b}}{r_{\rm A}} = \frac{R_{\rm b}}{R_{\rm A}}
\end{equation}
The orbital inclination, $i$, is another important photometric parameter. The relative depths of the primary and secondary eclipses yield the ratio of the surface brightnesses of the two stars, $J$, a quantity which is usually implicitly interpreted in terms of a \Teff\ ratio. Total eclipses allow very precise photometric parameters to be obtained. $k$ and $J$ can be strongly correlated when eclipses are not total, resulting in uncertain fractional radii and, in particular, a poorly constrained light ratio between the two stars. This problem can be solved by procuring an external measurement of the light ratio, such as from the spectral lines of the stars (e.g.\ \cite{MeClausen07}).

{\bf Eccentric orbits} present a complication by adding two more parameters to the mix: orbital eccentricity, $e$, and the longitude of periastron, $\omega$. These are best treated using the combination terms $e\cos\omega$ and $e\sin\omega$, which are much less strongly correlated than $e$ and $\omega$ themselves. In the approximation of small eccentricity and $i \sim 90^\circ$, the times and durations of the eclipses ($t_{\rm pri}$ and $t_{\rm sec}$, and $d_{\rm pri}$ and $d_{\rm sec}$) give the combination terms \cite{Kopal59}:
\begin{equation}
e\cos\omega \approx \frac{\pi}{2}\left(\frac{t_{\rm sec}-t_{\rm pri}}{P}-0.5\right)
\qquad \qquad \qquad \qquad
e\sin\omega \approx \frac{d_{\rm sec}-d_{\rm pri}}{d_{\rm sec}+d_{\rm pri}}
\end{equation}
The quantity $e\cos\omega$ is therefore dependent on the time of secondary minimum relative to primary minimum, which is easy to measure very precisely. Conversely, $e\sin\omega$ is given by the relative durations of the two eclipses, which is much less precise. Whilst photometry is good for constraining $e\cos\omega$ it is poor for specifying $e\sin\omega$. The opposite situation exists for radial velocities, so a well-behaved solution incorporates both types of observations.

{\bf Limb darkening} is a significant contributor to the shape of eclipses. Whilst a range of limb darkening laws exist\footnote{See \cite{Me08} and {\tt http://www.astro.keele.ac.uk/$\sim$jkt/codes/jktld.html}}, the linear law is satisfactory for the light curves of most EBs. The slightly more complex non-linear laws are important for studying the transits of extrasolar planets \cite{Me11} and perform similarly well to each other in practise. A wide range of theoretically-calculated coefficients for the limb darkening laws are available for observations taken in a standard photometric passband (e.g.\ \cite{Claret11}). When observations are of good quality, it may be better to fit for the coefficients of the laws rather than fixing them at theoretically-expected values (e.g. \cite{Me05}).

{\bf The extraction of photometric parameters from a light curve} requires a physical representation of the binary system. The `industry standard' computer program for this process is the Wilson-Devinney code \cite{Wilson71}, whose implementation of Roche geometry allows detached, semi-detached and contact binaries to be tackled. The input/output processes for this code are rather dated, making it an intimidating option for inexperienced researchers. An improved and extended version with a graphical user interface is now available, {\sc phoebe}\footnote{{\tt http://phoebe.fiz.uni-lj.si/}}, maintained by Andrej Pr\v{s}a \cite{Prsa05}. If one has a system whose stars suffer only mild tidal distortions and no mass transfer, the {\sc jktebop}\footnote{{\tt http://www.astro.keele.ac.uk/$\sim$jkt/codes/jktebop.html}} code is a good option. This code is a development of {\sc ebop} \cite{Etzel75,Etzel81}, maintained by the current author \cite{Me04}. Its chief advantages are simplicity and speed, which makes it well-suited to datasets which are large or require a high numerical precision. Another asset of speed is the ability to run extensive error analysis algorithms.

{\bf Correlations between photometric parameters} are inescapable, and lie in the binary stars' message to us rather than our methods of decoding the message. In such a situation it is important to perform careful error analyses which explore these correlations and return reliable error bars. Formal errors from a covariance matrix, as outputted by many analysis codes, are unreliable as they do not account well for parameter correlations. Alternatives include $\chi^2$-chase, Monte Carlo simulations \cite{Me04}, Markov Chain Monte Carlo \cite{Gregory05} and residual permutation \cite{Me08}. The {\it Numerical Recipes}\footnote{{\tt http://www.nr.com}} textbook \cite{Press92}, chapter 14, is an excellent starting point for interested readers.

\begin{figure} \centering
\includegraphics[width=1.0\textwidth,angle=0]{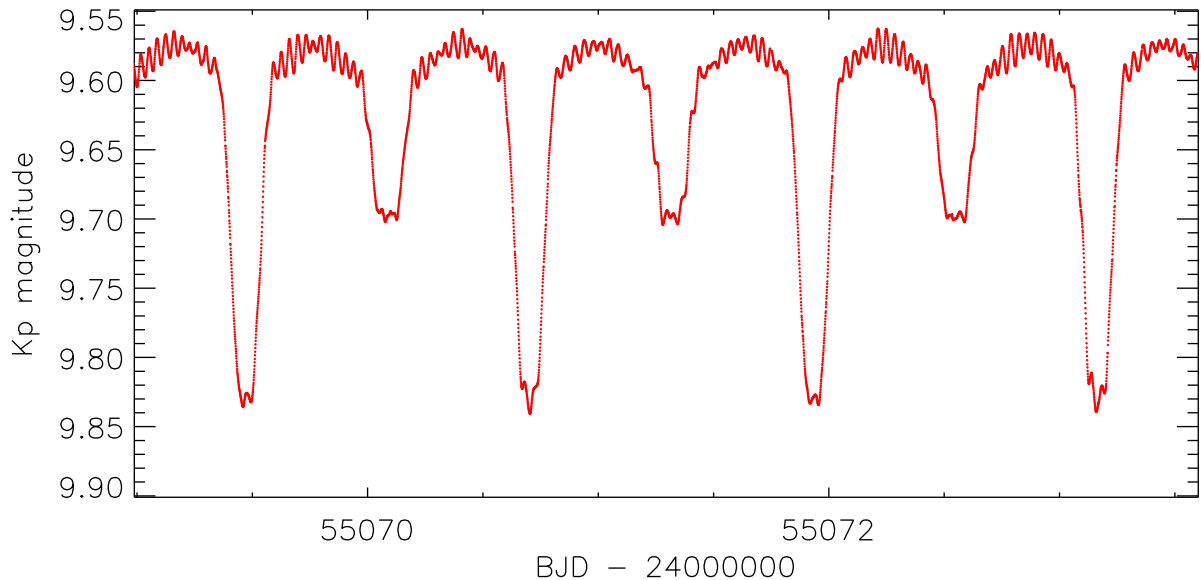}
\caption{{\it Kepler} satellite observations of the semi-detached
EB KIC\,10661783, whose primary star exhibits over 50 frequencies
arising from $\delta$\,Scuti pulsations \cite{Me1066}.} \end{figure}

\subsection{Radial velocity analysis of eclipsing binaries}

\noindent
Photometry is only half the story: a full analysis requires radial velocity (RV) measurements to measure the masses of the stars and the overall scale of the system. A typical requirement is thirty spectra of sufficient quality to reliably detect the spectral lines from both stars. They should be evenly distributed in orbital phase, but with a concentration on the quadrature phases when the velocities of the two stars are most divergent. Semi-detached EBs (such as Algol) are less demanding, as RVs are only needed for one of the stars. This is because the surface of the secondary star coincides with its Roche lobe, allowing its radius to be directly connected to the mass ratio.

For late-type stars, which rotate slowly, \'echelle spectrographs can provide marvellous results (e.g.\ \cite{Konacki09}). For early-type stars, which tend to have high rotational velocities and few spectral lines, moderate spectral resolution is sufficient but a high signal to noise ratio (S/N) is a necessity. For these, the mutual interference (blending) of lines from the two stars is a long-standing issue which could result in masses underestimated by 25\% \cite{Andersen75}.

RV measurements can be obtained in many ways: \\
{\bf Fitting Gaussian functions} to individual spectral lines (using a double-Gaussian for double lines) works well for isolated lines \cite{MeClausen07}.\\
{\bf Cross-correlation} \cite{Simkin74,Tonry79} allows one to include multiple spectral lines, but requires a template spectrum to compare to the observed spectra. It also suffers more from line blending, as contributions arise from the template as well as the observed spectra.\\
{\bf {\sc todcor} two-dimensional cross-correlation} \cite{Zucker94} uses different template spectra for the two stars. This is useful if the stars have quite different spectral characteristics, but is no better than normal cross-correlation if the stars are very similar \cite{MeClausen07}.\\
{\bf Broadening functions} \cite{Rucinski99} is a comparable approach to cross-correlation. This similarly requires a template spectrum but does not suffer from excess line blending due to this.\\
{\bf Spectral disentangling} \cite{Simon94} is a fundamentally different method which considers all observed spectra simultaneously and does not require a template spectrum. It assumes that each observed double-lined spectrum can be represented by the individual spectra of the two stars which have been shifted in RV and added together. This allows the best-fitting individual spectra to be calculated simultaneously with the orbital solution. Spectral disentangling does not suffer from line blending and returns high-quality spectra of the individual stars, but can be sensitive to continuum normalisation and may require a lot of CPU time. A detailed review is given by \cite{Pavlovski11}.

\subsection{Spectral analysis of eclipsing binaries}

\noindent
Whilst analysis of light and velocity curves is sufficient to give the masses and radii of the stars in an EB, they do not give luminosities or distance yet. For that we require effective temperature (\Teff) measurements. These can be obtained from calibrations against photometric colour indices (e.g.\ \cite{Casagrande06}) but this does not work well for early-type stars (partly due to the sensitivity to reddening) and requires colour indices for the individual stars. An alternative is to deduce spectral types and convert them to \Teff, once again using calibrations, but this suffers from the discreteness of the spectral classification system as well as uncertainties in the \Teff\ scale (particularly for high-mass and low-mass stars).

A better approach is to model the spectra of the stars directly, as is frequently done for single stars, in order to determine their atmospheric parameters. The details of the method vary widely depending on the type of star to be analysed, so I will not dwell on the methods here. Spectral disentangling holds an advantage for these analyses, as one can use the high-quality disentangled spectra rather than the individual composite spectra. EBs themselves show a major asset over single stars, in that their surface gravities are measure to such high accuracy that they can be fixed at the known values and thereafter ignored. This is particularly useful when measuring \Teff, which is often strongly correlated with surface gravity.

Once the surface gravity and \Teff\ of each star is known, its spectrum is well suited to deriving the chemical composition of its photosphere. This can be performed directly on the observed spectra in many cases (e.g.\ \cite{Clausen08}). Disentangled spectra are once again a useful alternative, due to their high S/N, but have a complication. The continuum light ratio of the stars cannot be determined in isolation, as such information simply does not exist in the observed spectra, so their continuum levels are unknown. This can be solved either by including the continuum level as an additional parameter to be  fitted for \cite{Tamajo11} or by using the light ratio(s) known from the photometric analysis to specify the continuum level directly (e.g.\ \cite{Pavlovski08}). Abundance analyses can then be carried out using the usual methods for single stars \cite{Pavlovski09}.

\subsection{Putting it all together}

\noindent
Analysis of the light curves generally gives: \ $P_{\rm orb}$, \ $r_{\rm A}$, \ $r_{\rm B}$, \ inclination $i$ \ and \ $e\cos\omega$ \\
Analysis of the RV curves normally gives: velocity amplitudes \ $K_{\rm A}$ and $K_{\rm B}$, \ plus $e\sin\omega$ \\
Then: \ \ $e\cos\omega$ \ and \ $e\sin\omega$ \ \ $\Rightarrow$ \ \ $e$ \ and \ $\omega$ \\
Then: \ \ $K_{\rm A}$ \ and \ $K_{\rm B}$ \ and \ $i$ \ and \ $e$ \ \ $\Rightarrow$ \ \ masses \ $M_{\rm A}$ \ and \ $M_{\rm B}$ \ and \ $a$ \\
Then: \ \ $r_{\rm A}$ \ and \ $r_{\rm B}$ \ and \ $a$ \ \ $\Rightarrow$ \ \ $R_{\rm A}$ \ and \ $R_{\rm B}$ \\
Then: \ \ $M_{\rm A}$ \ and \ $M_{\rm B}$ \ and \ $R_{\rm A}$ \ and \ $R_{\rm B}$ \ \ $\Rightarrow$ \ \ surface gravities and densities \\
Now we add in \Teff: \ \ ${\Teff}_{\rm A}$ \ and \ ${\Teff}_{\rm B}$ \ and \ $R_{\rm A}$ \ and \ $R_{\rm B}$ \ \ $\Rightarrow$ \ \ luminosities \ $L_{\rm A}$ \ and \ $L_{\rm B}$ \ and \ distance \\[3pt]
Distance measurement is an important result, and also relies on the existence of reliable apparent magnitudes in standard photometric passbands. Redder passbands are better (particularly near-infrared $JHK$) as they are less sensitive to interstellar extinction. The traditional way to measure distance is by using bolometric corrections to covert absolute bolometric magnitudes (which are calculated from the luminosities) to $V$-band absolute magnitudes and then relating these to the observed $V$-band apparent magnitude (see \cite{Me05iauc}). Alternative methods using surface brightness calibrations have been outlined by \cite{Lacy77} and \cite{Me05aa}, among others.

\begin{figure} \includegraphics[width=\textwidth,angle=0]{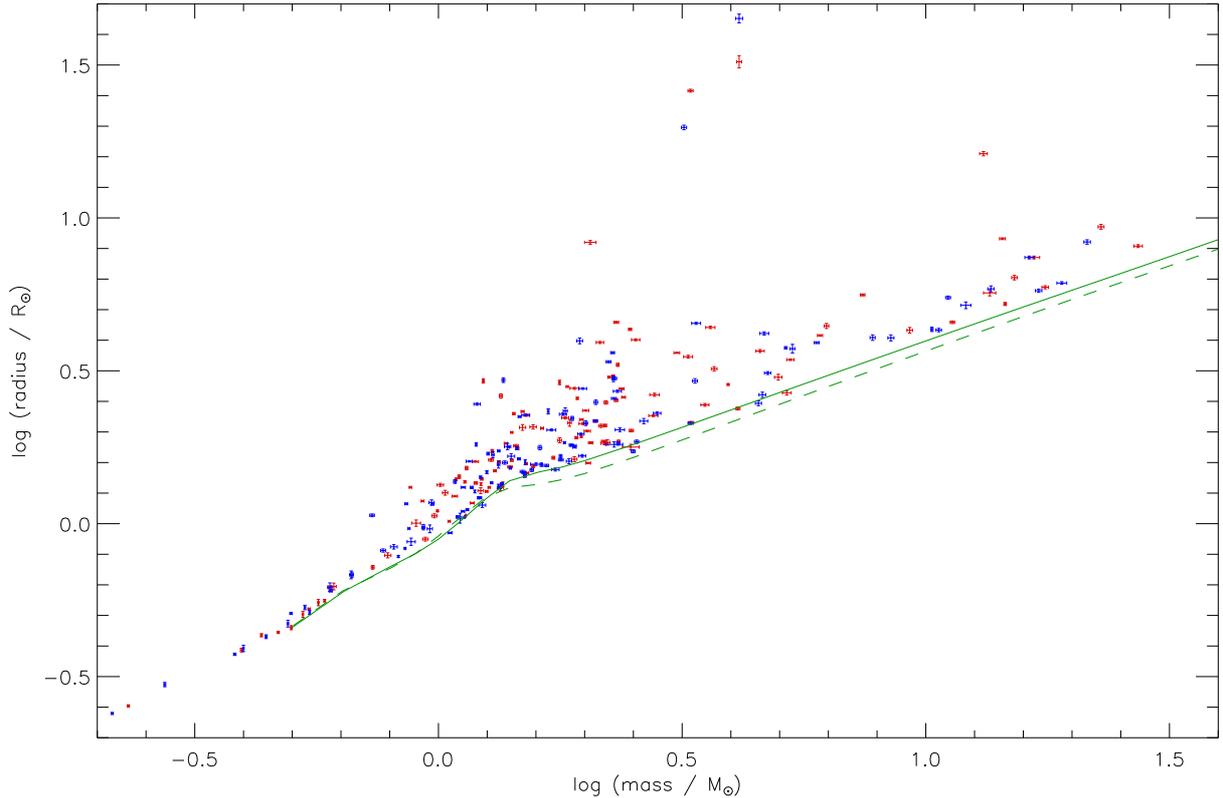}
\caption{Plot of the masses and radii of well-studied dEBs. The primary stars
are shown in red and the secondaries in blue. The green lines show the zero-age
main sequence relation from the Cambridge models for a solar metallicity
(unbroken line) and a half-solar metallicity (dotted line).} \end{figure}

\section{The Ghost of Christmas Yet to Come}

\noindent
The primary use of dEBs is as checks on the success of theoretical predictions (e.g.\ \cite{Pols97}) or to investigate the physical processes included in the models, such as mixing length \cite{Ludwig99}, convective core overshooting \cite{Claret07}, the chemical enrichment law \cite{Ribas99} and limb darkening coefficients from model atmospheres \cite{Claret08}. The study of dEBs in stellar clusters is a promising avenue \cite{Me04a} -- recent results for dEBs in the open cluster NGC\,6791 are of such high precision that the primary uncertainty in its age and chemical composition is now the variation between the predictions of different sets of stellar models \cite{Grundahl08,Brogaard11}. Models of low-mass stars have recently enjoyed renewed interest as vital ingredients in analyses of transiting extrasolar planets \cite{Me09}.

The cosmological distance scale rests on distances to nearby galaxies \cite{Freedman01} which are generally found using `standard candles' such as $\delta$\,Cepheids -- objects whose luminosities can be determined by calibrating them on nearby specimens. EBs can be direct indicators, rather than standard candles, because we do not need to calibrate them on neighbouring examples. Distance and reddening can instead be determined simultaneously using their measured luminosities (from \Teff\ and radius) and apparent magnitudes in multiple passbands. EBs have been used to measure distances to the nearby galaxies LMC \cite{Guinan98,Bonanos11}, SMC \cite{Hilditch05,North10}, M\,31 \cite{Vilardell10} and M33 \cite{Bonanos06}, as well as many open clusters within our own Milky Way.

The studies of EBs cited above are primarily distinguishable by the amount of effort required for each object: they can devour a lot of telescope time. Detailed analyses of EBs have now been available for roughly a century, during which only 127 objects have been studies in detail (a catalogue of well-studied dEBs [DEBCat] is maintained by the author at {\tt http://www.astro.keele.ac.uk/$\sim$jkt/debcat/}). By comparison, the {\it General Catalogue of Variable Stars} lists nearly 1000 `EA' systems, so supply already greatly exceeds demand.

Observational astronomy is currently undergoing an overwhelming shift towards survey projects, and in the future this will fundamentally change the opportunities to study EBs. Myriad large-scale photometric surveys are already being performed, and many more are planned. Most obtain repeated photometric measurements of point sources and so are able to discover new EBs. As an example, EB science has already been extracted from the ground-based planet search projects TrES and SuperWASP \cite{Devor08,Fernandez09,MeXY,Norton11}, and the similarly-targeted space missions CoRoT and {\it Kepler} \cite{Maceroni09,Derekas11,Me1066}. These missions hold the advantage of obtaining thousands of datapoints per object, often of extremely high quality, thus allowing important results to be obtained with modest or no reliance on follow-up observations. {\it Kepler} represents the current state-of-the-art for EB studies.

Further into the future, major missions such as GAIA and LSST will discover millions of EBs \cite{Tingley09,Prsa11} but obtain very few datapoints per object. A detailed study of each one is clearly hopeless. Automated methods will be able to deliver a limited number of parameters for many of these millions of EBs \cite{Devor08,Prsa08,Tingley09}, so the major advances will probably relate to statistical studies of larger populations of these objects. However, it will be possible to cherry-pick the most interesting or important EBs for further study. Areas meriting attention include: \\
{\bf Low-mass EBs}, as the properties of the known examples differ from theoretical predictions by up to 15\% in radius \cite{Lopez07} and low-mass stars host most of the known extrasolar planets. \\
{\bf High-mass EBs}, as massive stars are major contributors to the chemical and kinematic population of galaxies. \\
{\bf Asteroseismology of EBs}, which will allow pulsation frequencies to be measured for stars of known mass, radius, $T_{\rm eff}$ and photospheric chemical composition. \\
{\bf Giant-star EBs}, to provide constraints on stellar evolution \cite{Pietrzynski11}. \\
{\bf EBs in open clusters}, again for stellar evolutionary studies. \\
{\bf EBs in external galaxies}, for improving the local calibration of the cosmological distance scale.


\end{document}